\documentstyle[preprint,aps]{revtex}
\tightenlines
\begin{document}
\title{Financial Market Dynamics}
\author{Fredrick Michael and M.D. Johnson\cite{byline}}
\address{Department of Physics, University of Central Florida, Orlando,
FL 32816-2385}
\date{May 23, 2001}
\maketitle
\begin{abstract}
Distributions derived from non-extensive Tsallis statistics are
closely connected with dynamics described by a nonlinear
Fokker-Planck equation. The combination shows promise in
describing stochastic processes with power-law distributions and
superdiffusive dynamics.  We investigate intra-day price changes in
the S\&P500 stock index within this framework by direct analysis
and by simulation. We find that 
the power-law tails of the distributions, and the index's 
anomalously diffusing dynamics, are very accurately described 
by this approach. Our results show good
agreement between market data, Fokker-Planck dynamics, and simulation.
Thus the combination of the Tsallis non-extensive entropy and the
nonlinear Fokker-Planck equation unites in a very natural way the
power-law tails of the distributions and their superdiffusive dynamics.
\end{abstract}

\section{Introduction}
\label{intro}

In anomalously diffusing systems, a mean-square displacement 
scales with time according to a power law, $t^{\alpha}$,
with $\alpha>1$ (superdiffusion) or $\alpha<1$ (subdiffusion).
(The case $\alpha=1$ corresponds to normal diffusion.)
Anomalous diffusion has been observed in systems as widely varied as
plasma flow\cite{plasma}, surface growth\cite{surface}, and
financial markets\cite{econophysics}. 
A general framework for treating superdiffusive systems is provided
by the nonlinear Fokker-Planck equation, which is associated
with an underlying Ito-Langevin 
process\cite{plastino,tsallis-bukman,borland}.
This in turn has a very interesting connection to the
nonextensive entropy proposed by Tsallis\cite{tsallis}:
the nonlinear Fokker-Planck equation is solved by time-dependent
distributions which maximize the Tsallis entropy\cite{tsallis-bukman,zanette}.
This unexpected connection between thermostatistics and anomalous
diffusion gives an entirely new way to approach the study of
dynamics of anomalously diffusing systems. In this paper we
use this viewpoint to address the dynamics of financial markets.

Several financial markets' indices as well as their member stocks are
characterized by price changes whose variances have been shown to undergo
anomalous (super) diffusion under time
evolution\cite{econophysics,dacrrogna,mantegna1,mantegna}.
Moreover, the probability distributions of the price changes have power law
tails\cite{econophysics,mandlelbrot}. An open long-term question is how best to
describe these distributions and their time evolution. 
Earlier work has shown that the power-law tails can be described
by a distribution which maximizes the Tsallis entropy\cite{ramos,friedrich}.
The connection mentioned above then suggests that the market dynamics
might be controlled by a nonlinear Fokker-Planck equation.
In this description the power-law tails
and the anomalous diffusion arise together quite naturally.
Here we show that this approach seems to accurately
describe high-frequency
intra-day price changes for the S\&P500 index. It may also be suitable
to describe the superdiffusion and power law behaviors
observed in a broad range of markets and exchanges\cite{econophysics}.

The description we will use was developed in the general context of
anomalously diffusing systems by Tsallis and Bukman\cite{tsallis-bukman}
and Zanette\cite{zanette}.
Here we briefly summarize their results, which
are based on a maximization of entropy subject to certain constraints. 
The central ingredient is a time-dependent probability
distribution $P(x,t)$ of a stochastic variable $x$.  The
point of departure from normal maximum entropy approaches is the feature that
the entropy used is the non-extensive Tsallis entropy,
\begin{equation}
S_{q}=-{\frac{1}{1-q}}\left(  1-\int P(x,t)^{q}\,dx\right)  .\label{sq}%
\end{equation}
The Tsallis parameter $q$ characterizes the
non-extensivity of the entropy. 
In the limit $q\rightarrow1$ the entropy becomes the usual
logarithmic expression S=-$\int P\ln P$.

Associated with the non-extensive entropy is the use of the constraints
\begin{eqnarray}
&& \int P(x,t)\,dx=1,\label{norm}\\
&& \langle x-\bar{x}(t)\rangle_{q}\equiv\int[x-\bar{x}(t)]P(x,t)^{q}%
\,dx=0,\label{mean}\\
&& \langle(x-\bar{x}(t))^{2}\rangle_{q}\equiv\int[x-\bar{x}(t)]^{2}%
P(x,t)^{q}\,dx=\sigma_{q}(t)^{2}.\label{constraints}%
\end{eqnarray}
The first of these is simply the normalization of the probability. However, in
the latter two equations the probability distribution function is raised to
the power $q$. Unless $q=1$ these are not the usual constraints leading to the
mean and variance, and $\sigma_{q}^2$ (the `$q$-variance') is not the ordinary
variance.  Notice that the Tsallis nonextensivity
parameter $q$ is independent of time.

Maximizing the Tsallis entropy subject to these constraints, for fixed $q$,
yields\cite{tsallis-bukman}
\begin{equation}
P(x,t)={\frac{1}{Z(t)}}\left\{  1+\beta(t)(q-1)[x-\bar{x}(t)]^{2}\right\}
^{-\frac{1}{q-1}}.\label{pdf}
\end{equation}
Here parameters $Z$ (a normalization constant) and $\beta$ are
Lagrange multipliers associated with the first and third constraints,
and are given by
\begin{eqnarray}
Z(t) &=& {\frac{B\left(  {\frac{1}{2}},{\frac{1}{q-1}}-{\frac{1}{2}}\right)
}{\sqrt{(q-1)\beta(t)}}},\label{znorm}\\
\beta(t) &=& {\frac{1}{2\sigma_{q}(t)^{2} Z(t)^{q-1}}},\label{betaprime}
\end{eqnarray}
where $B(x,y)=\Gamma(x)\Gamma(y)/\Gamma(x+y)$ is Euler's Beta function. 

The ordinary variance of the distribution Eq.~(\ref{pdf}) is
\begin{equation}
\sigma^{2}(t)=\langle(x-\bar{x}(t))^{2}\rangle_1=\left\{
\begin{array}
[c]{cl}%
\frac{1}{(5-3q)\beta(t)} & ,\quad q<\frac{5}{3}\\
\infty & ,\quad q\geq\frac{5}{3}.
\end{array}
\right. \label{variance}%
\end{equation}
Hence for applications to data of finite variance, the Tsallis parameter $q$
must lie within the range
\begin{equation}
1\leq q<5/3.
\label{qrange}
\end{equation}
The Tsallis function Eq.~(\ref{pdf}) can be viewed as a {\it least biased}
probability distribution compatible with observed stochastic data with a
certain mean $\bar{x}(t)$ and variance $\sigma^2(t)$ (but with a
particular choice of $q$).

An important property of the probability distribution function Eq.~(\ref{pdf})
is that, with appropriate
time-dependent parameters, it is the solution of a time evolution equation
which leads naturally to anomalous diffusion\cite{tsallis-bukman,zanette}.
Consider the nonlinear Fokker-Planck equation
\begin{equation}
{\frac{\partial P(x,t)^{\mu}}{\partial t}}=-{\frac{\partial}{\partial x}%
}\left[  F(x)P(x,t)^{\mu}\right]  +\frac{D}{2}{\frac{\partial^{2}P(x,t)^{\nu}%
}{\partial x^{2}}},\label{fokker}%
\end{equation}
where $F(x)$ is a linear drift force, $F(x)=a-bx$. One can
show\cite{tsallis-bukman,zanette} that a
function of the form Eq.~(\ref{pdf}) solves this
as long as
\begin{equation}
q=1+\mu-\nu
\label{q}
\end{equation}
and the time dependences of the parameters are given by
\begin{eqnarray}
-{\frac{\mu}{\mu+\nu}}{\frac{dZ^{\mu+\nu}(t)}{dt}}&+&2\nu D\beta(t_0)Z(t_0)^{2\mu
}-bZ^{\mu+\nu}=0,\label{z}\\
{\frac{\beta(t)}{\beta(t_0)}}&=&\left(  {\frac{Z(t_0)}{Z(t)}}\right)^{2\mu},\label{beta}\\
{\frac{d\bar{x}}{dt}}&=&a-b\bar x.\label{xbar}%
\end{eqnarray}
By comparing
Eqs.~(\ref{znorm}) and (\ref{beta}) one easily shows that the nonlinear
Fokker-Planck equation preserves the norm ($\int P(x,t) dx$)
only if $\mu=1$. Hereafter we specialize to this case. 
Then Eqs.~(\ref{q},\ref{z},\ref{beta}) give
\widetext
\begin{equation}
\beta(t)^{-{3-q\over2}} = \beta(t_0)^{-{3-q\over2}}e^{-b(3-q)(t-t_0)}
-2Db^{-1}(2-q)
\left[ \beta(t_0)Z^2(t_0) \right]^{q-1\over2}
\left( e^{-b(3-q)(t-t_0)}-1 \right). \label{betat}
\end{equation}
$Z(t)$ is then obtained from Eq.~(\ref{znorm}).
\narrowtext

\section{Application to a Market Index}

It is apparent from the results sketched in the previous section that the
Tsallis probability distribution function $P(x,t)$ has properties which
make it a good candidate for describing the anomalous diffusion of
financial market indices, member stocks, and currency exchanges. 
First, in the large
$x$ limit $P(x,t)$ becomes a power law distribution,
\begin{equation}
P(x,t)\sim x^{-{\frac{2}{q-1}}}.\label{powerlaw}%
\end{equation}
This is in keeping with the observed power law tails in market distributions
and, indeed, certain market distributions have been well fit by
a distribution of the Tsallis form\cite{ramos,friedrich}. In fact this fit to
market data has a longer history --- the Tsallis function
is an extension to continuous values of $q$ of
the student-$t$ distribution, which has been known for
some time to provide a good fit to certain market data\cite{studentt,econophysics}.

What has not been widely recognized in the finance-related applications
of statistics is that the Tsallis distribution solves
the nonlinear Fokker-Planck equation. Consequently
it may provide a framework for understanding the dynamics of certain market
data, including anomalous diffusion. It is this possibility that we will 
now discuss.

For application to market indices the Tsallis parameter $q$ must
lie within the range Eq.~(\ref{qrange}), 
ensuring that the regular variance remains finite and evolves for
moderately long times as 
\begin{equation}
\sigma^{2}(t)\sim1/\beta(t)\sim t^{{\frac{2}{3-q}}}\label{sigmat}
\end{equation}
[using Eq.~(\ref{variance},\ref{betat}) in the case $b\ll 1$].
Thus the Fokker-Planck equation can describe superdiffusive processes
(with $1<q<5/3$).

Here we investigate one test case, the S\&P500 stock market index,
using fairly high-frequency 1-minute-interval data collected from
July 2000 to January 2001. 
This data consists of
a set of prices (index values) $p(\tau)$ at discrete times $\tau$
($\tau=1,2,\dots,50000$ trading minutes) obtained from the Terra-Lycos
QCharts server\cite{terra}. (The time period chosen has no special
significance.)
The quantity amenable to a stochastic analysis of the Ito-Langevin
type is the price change during various time 
intervals\cite{gardiner,friedrich,borland,stariolo}.
To agree with the notation used in the previous section we define 
$x$ to be a price change during a time interval $t$.
For each fixed time interval $t$ we generate a 
sequence of non-overlapping price changes $\{x_1,x_2,\dots\}$,
with $x_j=p(jt+1)-p((j-1)t+1)$.
We view the data thus generated for each fixed time interval $t$ as a sample
selected from a population with some distribution $P_{market}(x,t)$. 
Thus in this application the presumably stochastic variable $x$
is a price change, and the `time' variable $t$ is really the
corresponding time interval.

Anomalous diffusion occurs in random systems with correlations in time.
Because the price change during a longer interval is a sum of price
changes during shorter intervals, one has
\begin{equation}
P(x,t_1+t_2) = \int dx_1 P(x-x_1,t_2|x_1,t_1) P(x_1,t_1).
\end{equation}
where $P(\cdot|\cdot)$ is a conditional probability.
In the absence of correlation, price changes in different intervals
are independent, so that 
$P(x-x_1,t_2|x_1,t_1)=P(x-x_1,t_2)$. In this case it is easy
to see that the variance grows linearly with time. 
Anomalous diffusion thus means that price changes during successive
time intervals are not independent --- naturally enough, since
traders respond to earlier changes.  The observed superdiffusion of
financial markets thus indicates
correlation, and consequently a nontrivial time dependence.
Autocorrelation analyses of the S\&P500 market have found
strong correlations for times on the order of a few minutes, 
with weak persistent correlations at longer
times\cite{mantegna1,mantegna,econophysics}. 
We find the same for our data set. 
Consequently in this work we will concentrate on the intra-day market
dynamics of intervals less than one hour (1 min $\le t \le$ 60 min).

If the dynamics are of Fokker-Planck form, then there
must be a consistency between the distribution $P_{market}(x,t)$
at a certain time $t$ and the way the distributions evolve in time. 
One would have to find that the data at different
times can be fit by distributions of Tsallis form, with
a Tsallis parameter $q$ independent of time. If so, then
the variance will evolve in time according to
Eq.~(\ref{variance},\ref{betat}) with the same $q$ and appropriate
values for $D,b$.

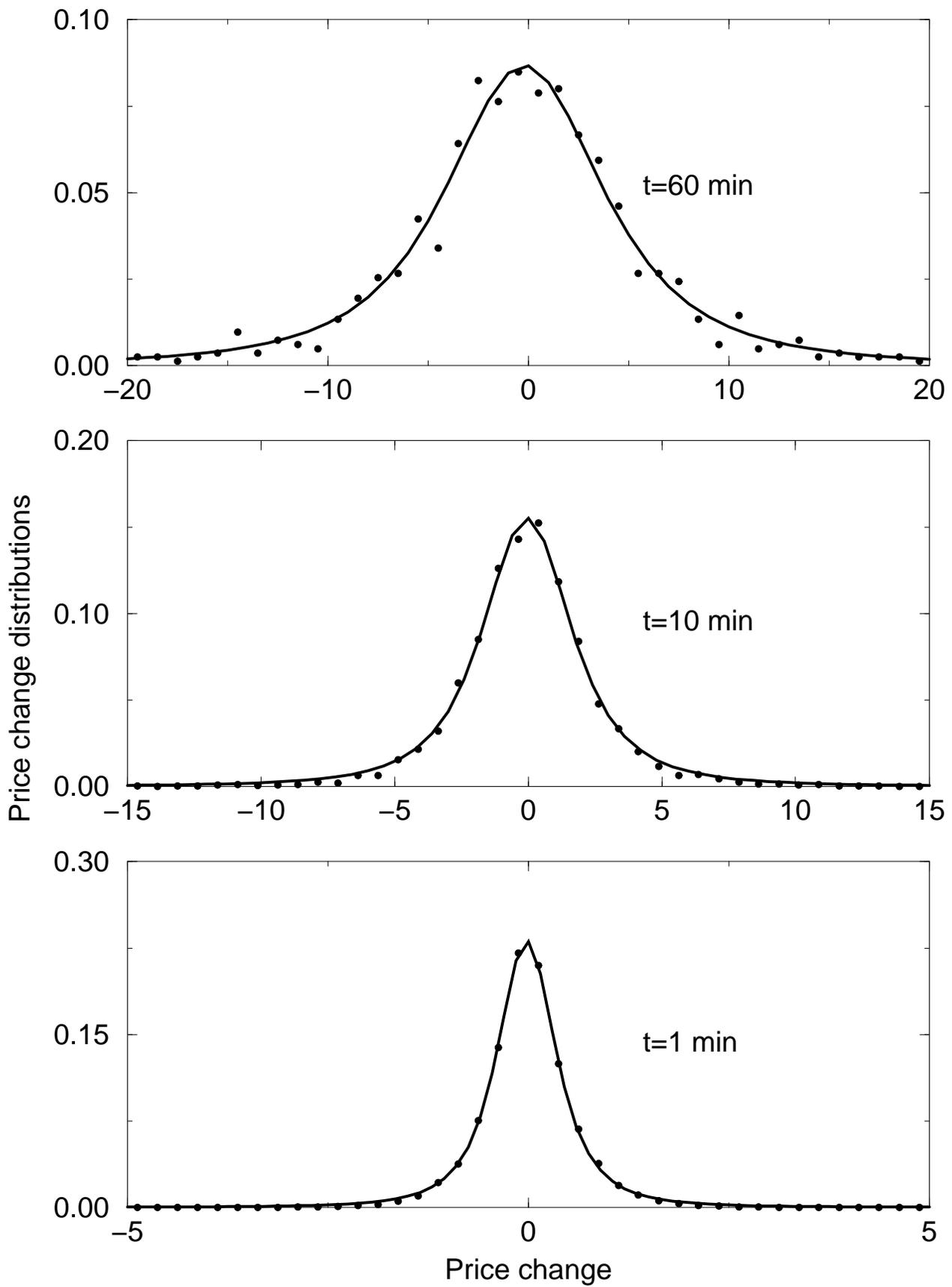
\begin{figure}
\caption{Distribution of S\&P500 index price changes at intervals
$t=$1, 10,  and 60 minutes, and a fit of Tsallis form [Eq.~(\ref{pdf})].
(Note: horizontal scales in the three plots are different.)}
\label{fit}
\end{figure}

We investigate the degree to which the S\&P500 price changes at different time
intervals $t$ can be described by a Tsallis distribution
evolving according to a Fokker-Planck equation.
The approach is to bin the data and perform a direct nonlinear 
$\chi^2$ fit of Eq.~(\ref{pdf}).
The `initial' distribution corresponds to the shortest value of $t$,
here $t_0=1$ minute. The $t_0=1\min$ price changes and a fit of Tsallis form
are shown in Fig.~\ref{fit}. The fit parameters are
$q=1.64\pm0.02$ and $\beta(t_0)=4.90\pm0.11$,
and consequently [from Eq.~(\ref{znorm})]
$Z(t_0)=1.09\pm0.02$.
We then fit the data at different time intervals $t$, with
the Tsallis parameter $q=1.64$ fixed and $\beta(t)$ determined by
the fit. It is important to this analysis that the data at all
times continues to be well fit by the Tsallis distribution, and indeed
this is what we find. (See Fig.~\ref{fit}.)

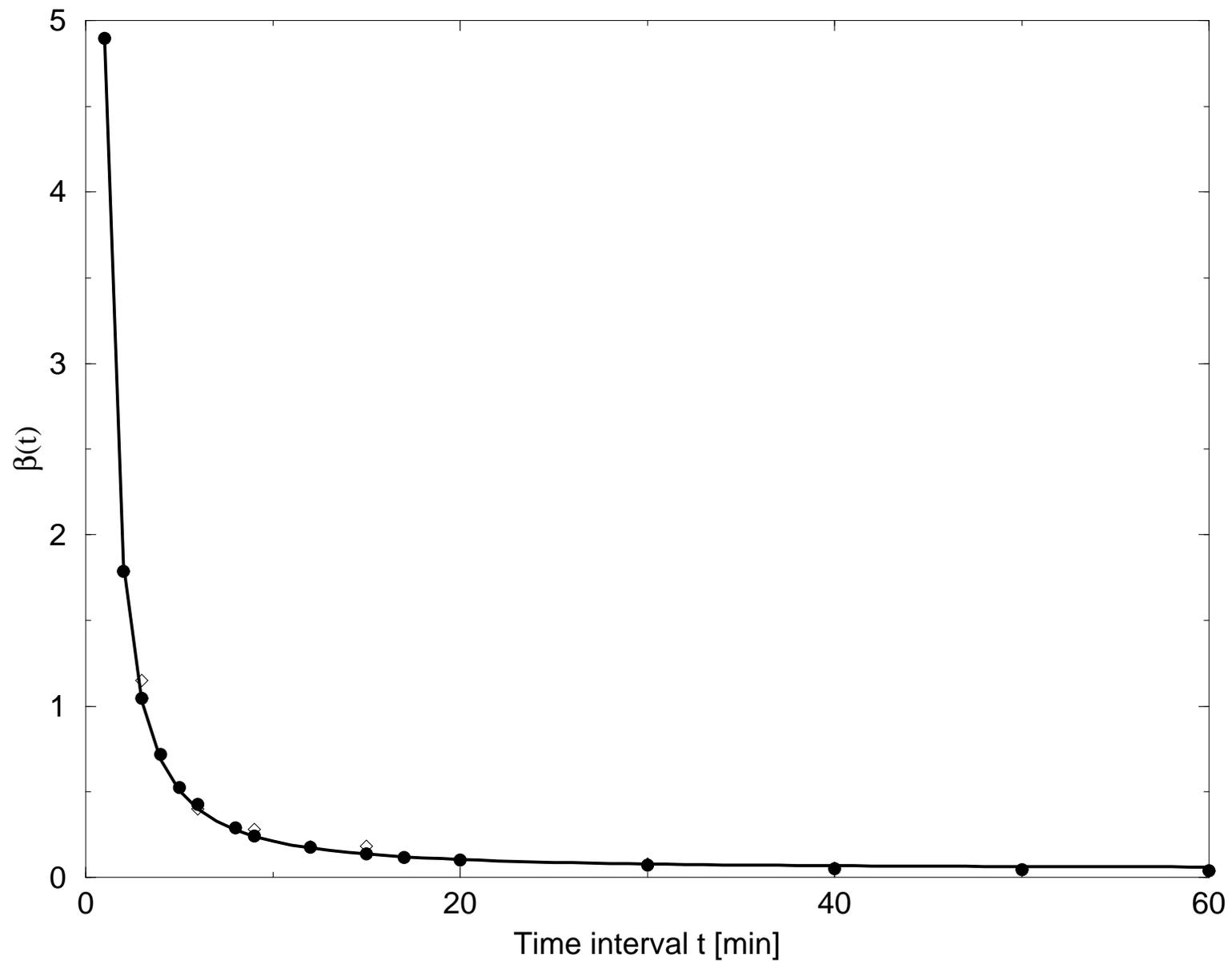
\begin{figure}
\caption{Parameter $\beta(t)$ (which is proportional to the inverse
of the variance) as a function of time interval $t$. Filled
circles: S\&P500 index data. Solid curve: Fit of the Fokker-Planck $\beta(t)$
from Eq.~(\ref{betat}).
Open diamonds: Simulated data using parameters from the
fit. Along most of the curve the simulated data are
indistinguishable from the actual market data.}
\label{betafig}
\end{figure}

The resulting inverse variance $\beta(t)$ extracted from the fits
is shown in Fig.~\ref{betafig}. Time evolution controlled by the
Fokker-Planck equation predicts the time dependence for $\beta(t)$
given in Eq.~(\ref{betat}). Accordingly we fit this form to the
extracted values, finding $D=0.217 \pm 0.003$ and $b=0.047\pm0.004$.
The Fokker-Planck form for $\beta(t)$ is also shown in Fig.~(\ref{betafig}).

It is clear from Fig.~\ref{betafig} that the agreement is quite good, indicating
time evolution of nonlinear Fokker-Planck form.

As a check on our analysis we also generated simulated data 
using a Tsallis distribution Eq.~(\ref{pdf}) evolving in time as described in 
Eqs.~(\ref{z},\ref{betat}). We then performed the same analysis that
we performed on the S\&P500 data. Briefly, the approach is as follows.
We use the parameters $q,D,b$ and initial values $\beta(t_0),Z(t_0)$
obtained from
the S\&P500 data. Parameters $\beta(t),Z(t)$ at
later times are obtained from Eq.~(\ref{betat},\ref{znorm}).
At several times $t$
we generate a set of random numbers $x$ with the Tsallis distribution
Eq.~(\ref{pdf}). This can be done by a transformation from a
uniformly distributed set of random numbers\cite{press}, as follows.
Consider a uniform distribution 
$P_{\mathrm uniform}(y)$ which is unity for $0\le y\le 1$ and zero otherwise.
By choosing an appropriate function $y(x)$ we can obtain the desired
distribution from the uniform distribution in the form
\begin{equation}
P(x,t)=P_{\mathrm uniform}(y)\left|  \frac{dy}{dx}\right|.\label{xform}
\end{equation}
This can be solved for $y$ as a hypergeometric function:
\begin{equation}
y=\int_{0}^{x}P(x^{\prime},t)dx^{\prime}=
x\,\cdot\, {}_{2}F_{1}\left[\frac{1}{2},\frac{1}{q-1},\frac{3}
{2},-(q-1)\beta(t) x^{2}\right].
\end{equation}
For each $y$ generated uniformly, we numerically invert this to find $x$.
The resulting $x$'s are distributed according to the Tsallis form, 
Eq.~(\ref{pdf}).
Thus the simulated price-change data is guaranteed to have a
Tsallis distribution evolving in time according to the Fokker-Planck
equation. We then analyzed this simulated data exactly as we did the
financial data as a check.  Fig.~\ref{betafig} shows that simulated data
closely tracks the actual market data.

\section{Conclusions}

The nonlinear Fokker-Planck equation for a probability distribution
$P(x,t)$ has a time-dependent solution
equivalent to the distribution obtained by maximization
of the Tsallis nonextensive entropy. Consequently
if a stochastic process has a Tsallis distribution it is natural to 
attempt to work backwards, and ask whether the underlying
dynamics are Fokker-Planck. 
We have investigated this possibility in one market, analyzing the
high-frequency intra-day dynamics of the S\&P500 index
within this framework.
We found that the distribution
$P(x,t)$ is well fit by the Tsallis form at all times, and evolves in time
according to the Fokker-Planck equation.  Thus the combination
of the Tsallis distribution and the Fokker-Planck equation
unites in a very natural way the power-law tails of the distributions and their
superdiffusive dynamics.

It would be of interest to apply this analysis to longer time windows,
inter-day (long term) dynamics, and also to other markets.

We acknowledge support from the NSF through grant DMR99-72683.

\end{document}